\documentclass[twocolumn,showpacs,pra,floatfix]{revtex4}

\usepackage{latexsym}
\usepackage{amsfonts}
\usepackage{amssymb}
\usepackage{dsfont}
\usepackage{mathrsfs}
\usepackage{psfrag}
\usepackage{graphicx}
\usepackage{amsmath}
\usepackage{bm}
\usepackage{psfrag}
\usepackage{color}
\usepackage{units}

\begin{document}

\title{Interplay of vacuum-mediated inter- and intraatomic couplings in a pair of atoms}

\author{Sandra Isabelle \surname{Schmid}}
\email{sandra.schmid@mpi-hd.mpg.de} 

\author{J\"org \surname{Evers}}
\email{joerg.evers@mpi-hd.mpg.de} 

\affiliation{Max-Planck-Institut f\"ur Kernphysik, Saupfercheckweg 1, D-69117
Heidelberg, Germany} 

\date{\today}

\newcommand{\lb}{\left\lbrace}
\newcommand{\rb}{\right\rbrace}
\newcommand{\g}[1]{\mf{\mc G}_{#1}}

\newcommand{\ket}[1]{\ensuremath{|#1\rangle}}
\newcommand{\bra}[1]{\ensuremath{\langle #1 |}}
\newcommand{\braket}[2]{\ensuremath{\langle #1 | #2 \rangle}}
\newcommand{\ov}[1]{\ensuremath{\overline{#1}}}
\newcommand{\mf}[1]{\boldsymbol{#1}}
\newcommand{\msz}[1]{\mathscr{#1}}
\newcommand{\mfi}[1]{\boldsymbol{#1}}
\newcommand{\ve}{\varepsilon}
\newcommand{\eps}{\epsilon}
\newcommand{\vro}{\varrho}
\newcommand{\tp}{t^\prime}
\newcommand{\be}{\begin{equation}}
\newcommand{\ee}{\end{equation}}
\newcommand{\ba}{\begin{eqnarray}}
\newcommand{\ea}{\end{eqnarray}}
\newcommand{\bali}{\begin{align}}
\newcommand{\eali}{\end{align}}
\newcommand{\vfi}{\varphi}
\newcommand{\mc}[1]{\ensuremath{\mathcal{#1}}}
\newcommand{\ix}[1]{\text{ #1}}
\newcommand{\bc}{\begin{center}}
\newcommand{\ec}{\end{center}}
\newcommand{\bi}{\begin{itemize}}
\newcommand{\ei}{\end{itemize}}
\newcommand{\mean}[1]{\ensuremath{ \langle\,#1\, \rangle}}
\newcommand{\means}[1]{\ensuremath{ \langle\,#1\, \rangle_{\text{st}}}}
\newcommand{\meanb}[1]{\ensuremath{ \big\langle\,#1\, \big\rangle}}
\newcommand{\meansb}[1]{\ensuremath{ \big\langle\,#1\, \big\rangle_{\text{st}}}}
\newcommand{\meanF}[1]{\ensuremath{ \text{Tr} [#1\, \rhoF ]}}
\newcommand{\meanI}[1]{\ensuremath{ \langle #1  \rangle_0}}
\newcommand{\meanbI}[1]{\ensuremath{ \big \langle #1 \big\rangle_0}}

\newcommand{\rf}{\vro_{\text{F}}}
\newcommand{\gwtro}{\widetilde{\mf\vro}_{at,tot}}
\newcommand{\wtro}{\widetilde{\mf\vro}_{at}}
\newcommand{\wto}{\widetilde{\vro}_{at}}

\newcommand{\dwtro}{\widetilde{\mf\vro}_{D}}
\newcommand{\ups}{\upsilon}

\newcommand{\wtst}{\widetilde{\mf\vro}_{\text{st}}}
\newcommand{\rft}{\tilde{\vro}_{\text{F}}}
\newcommand{\sa}{s}
\newcommand{\di}{d}

\newcommand{\Sp}[2]{\mathcal{S}_{#1 \,+}^{\,#2}}
\newcommand{\Sm}[2]{\mathcal{S}_{#1 \,-}^{\,#2}}
\newcommand{\Spm}[2]{\mathcal{S}_{#1 \,\pm}^{\,#2}}
\newcommand{\sSp}[2]{\tilde{\mathcal{S}}_{#1 \,+}^{\,#2}}
\newcommand{\sSm}[2]{\tilde{\mathcal{S}}_{#1 \,-}^{\,#2}}
\newcommand{\sSpm}[2]{\tilde{\mathcal{S}}_{#1 \,\pm}^{\,#2}}
\newcommand{\wt}[1]{\widetilde{#1}}

\newcommand{\Gpm}[2]{\mf{G}_{#1 \,\pm}^{\,#2}}
\newcommand{\Gm}[2]{\mf{G}_{#1 \,-}^{\,#2}}
\newcommand{\Gp}[2]{\mf{G}_{#1 \,+}^{\,#2}}

\newcommand{\cross}[2]{\mf{\zeta}(#1, #2)}
\newcommand{\crosscon}[2]{\mf{\zeta^*}(#1, #2)}
\newcommand{\bcross}[2]{\mf{\zeta}_B(#1, #2)}
\newcommand{\bcrosscon}[2]{\mf{\zeta}_B^*(#1, #2)}

\newcommand{\spec}[5]{\mathcal{T}_{\,{#1}{{#2}}}^{\,{#3}{#4}}(#5)}
\newcommand{\pigeo}[2]{\,e^{ik_0\mf{\hat{R}}_\pi(\mf{r}_#1-\mf{r}_#2)}}
\newcommand{\sigeo}[2]{\,e^{ik_0\mf{\hat{R}}_\sigma(\mf{r}_#1-\mf{r}_#2)}}
\newcommand{\sgeo}[2]{\,e^{ik_s\mf{\hat{R}}(\mf{r}_#1-\mf{r}_#2)}}
\newcommand{\sdgeo}[2]{\,e^{ik_{s/d}\mf{\hat{R}}(\mf{r}_#1-\mf{r}_#2)}}
\newcommand{\geo}[2]{\,e^{ik_0\mf{\hat{R}}(\mf{r}_#1-\mf{r}_#2)}}
\newcommand{\rhat}{\mf{\hat R}}

\newcommand{\I}[2]{\mathcal{I}_{\text{#1}}^{ \,#2}}
\newcommand{\dip}[1]{\mf{\hat{d}}^{\,#1}}
\newcommand{\dm}[2]{\mf{d}_{#1}^{\,#2}}
\newcommand{\near}{\Omega_N}
\newcommand{\far}{\Omega_F}
\newcommand{\TensorRe}{ \overset{\leftrightarrow}{\chi}_{\text{re}}}
\newcommand{\TensorIm}{ \overset{\leftrightarrow}{\chi}_{\text{im}}}
\newcommand{\rot}[3][]{\mc{R}_{#2}^{#1}(#3)}
\newcommand{\rotop}[3][]{\hat{R}_{#2}^{#1}(#3)}

\newcommand{\rhoA}{\vro_{\text{A}}}
\newcommand{\rhoT}{\tilde{\vro}_{\text{A}}}
\newcommand{\rhoF}{\vro_{\text{F}}}
\newcommand{\rhost}{\vro_{\,\text{st}}}
\newcommand{\MM}{\mf{\mc{M}}}

\newcommand{\del}[1]{\partial_{#1}}

\begin{abstract}
The resonance fluorescence emitted by a system of two dipole-dipole interacting nearby four-level atoms in $J=1/2 \leftrightarrow J=1/2$ configuration is studied. This setup is the simplest realistic model system which provides a complete description of the (interatomic) dipole-dipole interaction for arbitrary orientation of the interatomic distance vector, and at the same time allows for intraatomic spontaneously generated coherences. 
We discuss different methods to analyze the contribution of the various vacuum-induced coupling constants to the total resonance fluorescence spectrum. These allow us to find a dressed state interpretation of the contribution of the different interatomic dipole-dipole couplings to the total spectrum. We further study the role of the spontaneously generated coherences, and identify two different contributions to the single-particle vacuum-induced couplings. We show that they have a noticeable impact on the total resonance fluorescence spectrum down to  small interatomic distances, even though the dipole-dipole couplings constants then are much larger in magnitude than the the single-particle coupling constants. Interestingly, we find that the interatomic couplings can induce an effect of the intraatomic spontaneously generated coherences on the observed spectra which is not present in single-atom systems.
\end{abstract}

\pacs{42.50.Nn, 42.50.Lc, 42.50.Gy}


\maketitle

\allowdisplaybreaks

\section{Introduction}

\begin{figure}[t]
\centering
\includegraphics[width=8.5cm]{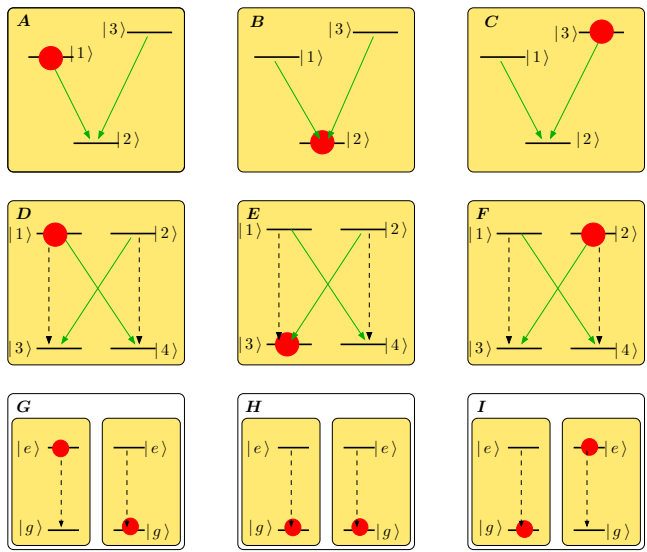}
\caption{\label{SGC}Vacuum-mediated couplings between different transition dipole moments. (A)-(C) illustrate intraatomic or single-particle vacuum-induced (SPVC) couplings between different transition dipole moments in a single atom. (D)-(F) show a generalized form of these SPVC in a four-level $J=1/2\leftrightarrow J=1/2$ level scheme. (G)-(I) depict interatomic or two-particle vacuum induced (TPVC) couplings between transition dipole moments in different atoms. In all three cases, the system is initially excited on one of the involved transitions, see (A,D,G). In the second step, a virtual photon is emitted together with the de-excitation of the atom (B,E,H). Finally, the virtual photon is reabsorbed on the second transition (C,F,I). }
\end{figure}

Coherence and interference effects form the basis of many quantum mechanical phenomena~\cite{book-ficek,zubairy:qo}. Its applications have been revolutionized by the invention of the laser as a coherent source of light. But somewhat surprising, under certain conditions it is also possible for coherences to be created in the interaction with the vacuum.
An archetype system in which these {\it spontaneously generated coherences} (SGC) have been predicted is the three-level $V$-type system shown in Fig.~\ref{SGC}(A-C)~\cite{book-agarwal,agassi,syzhu}. The relevant physical processes can be understood intuitively as follows. Suppose the atom is initially in state $|e_1\rangle$. The atom can be de-excited to the ground state $|g\rangle$ by the emission of a photon into the vacuum. This photon can leave the system, giving rise to spontaneous emission. Alternatively, it can be reabsorbed on the same transition, which leads to the Lamb shift. Finally, it could be reabsorbed on the second transition, with final state $|e_2\rangle$. The latter process leads to the creation of SGC between the two excited states, and it has been shown in many theoretical works that such SGC could give rise to fascinating applications~\cite{book-ficek,v1,v2,v3,v4,v5,v6,v7,v8,v9,v10,v11,v12,v13}. The  interpretation is facilitated by  a quantized treatment of the light fields~\cite{quantized}. In contrast, so far there are no conclusive observations of SGC in atomic systems~\cite{exp1}. The reason for this is that the process leading to SGC only occurs if both the two involved transitions are near-degenerate, and if their dipole moments are non-orthogonal. Speaking pictorially, the photon then cannot distinguish between the two transitions. Unfortunately, these conditions usually are not fulfilled in real atoms. A proof-of-principle experiment verifying the presence of SGC could however be achieved in quantum dots~\cite{dot}.

Recently it was found that there is a variant of such vacuum-mediated couplings with slightly relaxed conditions~\cite{time-energy,martin1,agarwal-inc}. In this case, the two transitions involved do not share a common state, as it is the case e.g. with $|g\rangle$ in the $V$-type scheme. The simplest example is the four-level $J=1/2 \leftrightarrow J=1/2$ setup, see Fig.~\ref{SGC}(D-F). One might be tempted to conclude that vacuum-mediated interactions between the two transitions are not possible, because two ground states of the transitions are orthogonal, $\langle S_1^-\, S_2^+ \rangle = 0$ with $S_1^- = |3\rangle\langle 1|$ and $S_2^+= |4\rangle\langle 2|$. But it turns out that there are observables which nevertheless are affected by SGC. An example is the resonance fluorescence spectrum, which depends on $\langle S_1^-(t)S_2^+(t+\tau)\rangle $. This two-time correlation was shown to be non-zero in general, which can be understood from the fact that the atom may evolve between the two ground states $|3\rangle$ and $|4\rangle$ in the time delay $\tau$. It is therefore possible to observe effects of SGC in realistic atomic systems. In the following, we will denote couplings of this type as {\it intra-atomic couplings} or {\it single particle vacuum couplings} (SPVC), since two transitions within the same atom are coupled.

Similar vacuum-mediated photon exchange processes can also take place between two transition dipoles belonging to different atoms~\cite{dicke,Le1970a,thirunamachandran,FiTa2002,Fi1991,AgPa2001,e1,e2,e3,e4,e5,geometry,twotwolevel,breakdown,dfs,sandra}. Fig.~\ref{SGC}(G-I) illustrates this for two two-level systems. Here, one atom is de-excited and emits a virtual photon, and the second atom absorbs this photon and is excited from the ground state to the excited state.
Processes of this type are known as dipole-dipole couplings, and will be called {\it inter-atomic couplings} of {\it two-particle vacuum couplings} (TPVC) in the following. The TPVC  crucially influence the dynamics of the combined system, as can be seen, for example, from the resonance fluorescence spectrum~\cite{e5,twotwolevel}. Such an energy transfer process between two particles is only possible, if the distance $r$ separating the two atoms is small compared to the respective transition wavelength, and if the two transitions are near-degenerate. In contrast to the SPVC, however, in general for TPVC there is no restriction on the dipole moments of the two transitions~\cite{AgPa2001,geometry,twotwolevel,breakdown,dfs,sandra}. The TPVC between orthogonal dipole moments only vanish in certain relative alignments of the interatomic distance vector and the involved dipole moments. Therefore, in general TPVC have to be described using complete angular momentum multipletts in order to obtain correct results~\cite{breakdown}.

\begin{figure}[t]
\centering
\includegraphics[width=\columnwidth]{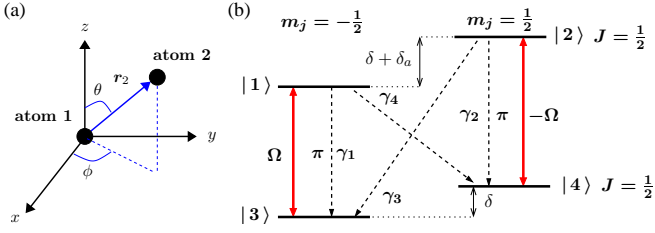}
\caption{\label{system}The analyzed model system consisting of two nearby four-level atoms in $J=1/2\leftrightarrow J=1/2$ atoms. This configuration is the simplest realistic system in which both single-particle and two-particle vacuum-induced couplings occur. In (a) the geometrical setup is shown. We assume that atom 1 is located at the origin and atom 2 at the point $\mf r_2$. (b) shows the inner structure of each of our two atoms in $J=1/2\rightarrow J=1/2$ configuration. Note that the energy differences are not to scale.}
\end{figure}

From the above discussion it is clear that TPVC and SPVC are closely connected, and this is also reflected in their similar theoretical description. Motivated by this and by the many applications that have been suggested for TPCV and SPVC individually, here we study the simplest realistic atomic system in which both TPVC and SPVC can occur. This system consists of two dipole-dipole interacting four-level systems in $J=1/2\leftrightarrow J=1/2$ configuration, see Fig.~\ref{system}. In this setup, each individual atom is modelled by a complete set of angular momentum states, and fulfills the conditions for the generalized SPVC. In addition, the four dipole-allowed transitions in each atom interact with the corresponding four transitions in the second atom, giving rise to TPVC both between parallel and between orthogonal transition dipole moments. Our general aim is to study the impact of the couplings of the system dynamics and on the optical properties. More specifically, we are interested in the role of the SPVC, and the dependence of the SPVC contribution on the interparticle couplings. As main observable, we discuss the resonance fluorescence spectrum of the two-atom system. In order to understand the  effects of the various couplings, we analyze different interparticle distance classes. In the small-distance limit, the TPVC coupling coefficients are much larger than all other relevant system parameters. In the large-distance case, the TPVC vanish, such that the SPVC are the dominant vacuum couplings. In an intermediate case, both TPVC and SPVC coupling parameters are of similar order. We interpret our results using two different methods. The first analysis is based on the eigenvalues of the matrix governing the system dynamics, which describe position and width of the different eigenstates of the system. Secondly, by artificially switching individual couplings on or off in the analysis, the quantitative impact on the results can be studied. A combination of both methods allows to understand the formation of multiply dressed states in the system in detail. Regarding the SPVC, we find that the contributions to the spectrum by the single particle vacuum-induced couplings survive even at very low interparticle distance, where the TPVC coupling coefficients are much larger than the corresponding SPVC coefficients. Therefore, also in the two-particle case,  the $J=1/2\leftrightarrow J=1/2$ configuration enables one to observe spontaneously generated coherences in a realistic level scheme. We distinguish two different types of intraatomic couplings, which contribute either directly in the equations of motion, or in the total expression for the resonance fluorescence. We find that in particular for small interatomic distances, the parts entering the expression for the spectrum are the dominating intraatomic coupling contributions, as in the single-particle case. But in contrast to the single-particle case, also the SPVC contribution in the equations of motion can significantly contribute to the obtained fluorescence spectra in the two-atom case. We thus conclude that the TPVC can induce additional SPVC.

The outline of the paper is as follows. In Sec.~\ref{sec-model} we provide the theoretical background for our analysis. We start by deriving the equations of motion for our biatomic system. We then derive expressions for the resonance fluorescence spectrum as our main observable. Finally, we present the eigenstate spectrum of our system as a tool to analyze the contribution of the different coupling constants. Sec.~\ref{sec-result} contains our results. We start by discussing the eigenstate spectrum in Sec.~\ref{sec-eigenstate}, and then proceed to discuss the role of interatomic couplings in Sec.~\ref{sec-inter}. Finally, in Sec.~\ref{sec-intra}, we analyze intraatomic couplings. Sec.~\ref{sec-summary} concludes with a summary.

\section{\label{sec-theory}Theoretical analysis}
\subsection{\label{sec-model}Model system}

We consider a system consisting of $N=2$ atoms in $J=\frac{1}{2}\leftrightarrow J=\frac{1}{2}$ configuration. Each atom has $L=4$ levels and $D=4$ possible electric dipole allowed (E1) transitions with dipole moments $\dm{i}{}$ ($i\in\{1,..,4\}$) as shown in Fig.~\ref{system}. $\dm{1}{}$ and $\dm{2}{}$ couple to linearly polarized light, the so-called $\pi$ transitions, whereas the $\sigma$ transitions $\dm{3}{}$ and $\dm{4}{}$ couple to circularly polarized photons. We define the mean transition frequency as
\be 
\omega_0=\frac{1}{D}\sum_{i=1}^D\omega_i\,,
\ee
where $\omega_i$ denotes the transition frequency of the $i$-th  transition.
Calculating the dipole matrix elements for our system via the Wigner-Eckart theorem \cite{sakurai} we find
\ba 
\dm{1}{}=\bra{1}_\mu\dip{}\ket{3}_\mu&=-\frac{1}{\sqrt{3}}\mc D\,\mf e_z\:,\hspace{0.2cm} \dm{2}{}&=-\dm{1}{}\nonumber\\
\dm{3}{}=\bra{2}_\mu\dip{}\ket{3}_\mu&=\sqrt{\frac{2}{3}}\mc D\,\mf e_-\:,\hspace{0.2cm} 
\dm{4}{}&=(\dm{3}{})^{\,*}\notag\:.
\label{d1}
\ea 
Here, $\mc D$ is the reduced dipole matrix element, $\mf e_-=(\mf e_x-i\mf e_y)/{\sqrt 2}$ the circular polarization vector, $\mu\in\{1,2\}$, and $\mf e_x, \mf e_y, \mf e_z$ are the Cartesian unit basis vectors.
We assume the system to be driven by a monochromatic laser beam propagating in the $y$ direction,
\be 
\mf E_L(t)=\mc E_L\,e^{i (\mf k_L \mf r -\omega_Lt)}\, {\mf \epsilon} + \text{c.c.}\:.
\ee
$\mc E_L$ is the field amplitude, $\mf k_L$ the wave vector, $\omega_L$ the frequency, and we choose the polarization ${\mf \epsilon} = {\mf e}_z$ such that the driving field only couples to the $\pi$-transition dipoles $\dm{1}{}$ and $\dm{2}{}$.

The free evolution of our two atoms is governed by the Hamiltonian
\be 
{\mc H}_{at}=\hbar\sum_{\mu=1}^{2}\sum_{i=1}^{2}\left(\omega_i\,\Sp{i}{\mu}\Sm{i}{\mu}+
\omega_{i+2}\,\Sm{i}{\mu}\Sp{i}{\mu}\right)\:,
\label{has}
\ee
with $\Sp{i}{\mu}$ the atomic excitation operator for transition $i$ in atom $\mu$, and $\Sm{i}{\mu}$ is the corresponding de-excitation operator.
In Schr\"odinger\textquoteright s picture the interaction with the laser field is described by
\be 
\mc H_L=-\hbar\sum_{\mu=1}^2\sum_{i=1}^4\left(\Omega_i(\mf r_\mu)\,e^{-i\omega_Lt}\Sp{i}{\mu}+\ix{H.c.}\right)\:.
\label{hls}
\ee
Here, the position dependent Rabi frequency is defined as
\begin{subequations}
\label{rabi}
\ba 
\Omega_i(\mf r)&=&\Omega_i\,e^{i\mf k_L\cdot \mf r}\:, \\
\Omega_i&=&\frac{\mf d_i\cdot \mf\epsilon \: \mc E_L}{\hbar}\,.
\ea
\end{subequations}
Since $\dm{1}{}$ and $\dm{2}{}$ are antiparallel, we can define
\be
\Omega(\mf r)=\Omega\,e^{i\mf k_L\cdot \mf r} = \Omega_1(\mf r)=-\Omega_2(\mf r)\,.
\ee
In a suitable interaction picture with $\Delta_i=\omega_L-\omega_i$ the laser detunings, the full Hamiltonian then reads
\be 
\tilde{\mc H}=-\hbar\sum_{\mu=1}^{2}\sum_{i=1}^{2}\left[\Delta_i\sSp{i}{\mu}\sSm{i}{\mu} + \left(\Omega_i(\mf r_\mu)\sSp{i}{\mu}+\ix{H.c.}\right)\right]\,.
\ee
Finally, the system dynamics can be described by the master equation
\be 
\del{t}\tilde\vro_{at}(t)=\frac{1}{i\hbar}[\tilde{\mc H},\tilde\vro_{at}(t)]+\mc L_\Omega\tilde\vro_{at}(t)+\mc L_\gamma\tilde\vro_{at}(t)\:.
\label{Hg}
\ee 
Here,
\be
\mc L_\Omega\tilde\vro_{at}(t)=i  \sum\limits_{\genfrac{}{}{0pt}{2}{\mu,\nu=1}{\mu\not=\nu}}^N 
 \sum\limits_{i=1}^{D}\sum\limits_{j=1}^{D} 
 \Omega_{ij}^{\mu\nu} \left[ \sSp{i}{\mu}\sSm{j}{\nu}, \wto(t) \right]
\ee
is a modification to the coherent part of the evolution arising from the dipole-dipole coupling (TPVC) of the two atoms, as can be seen from the restriction $\mu\neq \nu$ of the summation. It leads to the formation of collective dressed states, similar to the well-known symmetric and anti-symmetric collective states in two interacting two-level atoms~\cite{book-ficek}. It can be seen that couplings between all four transitions in one of the atoms to all four transitions in the second atom are considered.
The incoherent part is given by
\begin{align}
&\mc L_\gamma\tilde\vro_{at}(t)=-\sum\limits_{\mu,\nu =1}^N  
 \sum\limits_{i=1}^{D}\sum\limits_{j=1}^{D}  
  \Gamma_{ij}^{\mu\nu}
\left( \sSp{i}{\mu}\sSm{j}{\nu}\wto(t)\right.\nonumber\\ 
&+ \left. \wto(t) \sSp{i}{\mu}\sSm{j}{\nu}
-2 \sSm{j}{\nu}  \wto(t) \sSp{i}{\mu}\right)\:.
\end{align}
This term essentially contains three types of contributions. For $\mu=\nu$ and $i=j$, i.e., absorbing and emitting transitions are identical, the term describes the usual spontaneous emission with the rates $\gamma_i$ of the individual transitions $i$, 
\be
\gamma_i = \Gamma_{ii}^{\mu\nu}\,.
\ee
Second, if the energy exchange occurs between different transitions in the same atom ($\mu=\nu$, but $i\neq j$), then the corresponding process is a SPVC governed by the coupling constant 
\be 
\Gamma_{ij}^{\mu\mu}=\sqrt{\gamma_i\gamma_j}\: \frac{\mf d_i\cdot\mf d_j^*} {|\mf d_i||\mf d_j|}\,.
\ee
Here, the normalized scalar product of the two dipole moments accounts for the fact that SPVC only occur between non-orthogonal transition dipole moments.
Finally, the terms with  $\mu \neq \nu$ and $i \neq j$ describe TPVC with coupling constants $\Gamma_{ij}^{\mu\nu}$. The incoherent and coherent TPVC coupling constants van be expressed as 
\be 
\Gamma_{ij}^{\mu\nu}=\frac{1}{\hbar}\left[(\dm{i}{})^\ix T \text{Im}\overset{\leftrightarrow}{\chi}(\mf r_{\mu\nu}) \dm{i}{*}\right]
\ee
and
\be 
\Omega_{ij}^{\mu\nu}=\frac{1}{\hbar}\left[(\dm{i}{})^\ix T \text{Re}\overset{\leftrightarrow}{\chi}(\mf r_{\mu\nu}) \dm{i}{*}\right]\:
\ee
with the help of the tensor
\begin{align}
\overset{\leftrightarrow}{\chi}_{pq}(\mf{r})=&
\frac{(k)^3}{4\pi\ve_0}\left [ \delta_{pq} \left (  
\frac{1}{\eta}
+\frac{i  }{\eta^2} - \frac{1}{\eta^3} \right ) \right.\notag\\
 &- \left. \frac{[\mf{r}]_{p}[\mf{r}]_{q}}{r^2} \left( \frac{1}{\eta} + \frac{3i}{\eta^2} 
- \frac{3}{\eta^3} \right )\right ]\,e^{i \eta}\:.\notag
\label{chi}
\end{align}
In the following we assume the energy differences $\delta=\delta_a=0$. This corresponds to the case where no external magnetic field is applied. Then the laser detunings $\Delta_1$ and $\Delta_2$ become equal and we denote them by $\Delta$. Additionally, all transition frequencies become equal  $\omega_i=\omega_0$.

\subsection{\label{subspec}Resonance fluorescence spectrum}
The total resonance fluorescence spectrum is given by the Fourier transform of 
the two-time correlation function of the electric field operators~\cite{zubairy:qo}
\be
\text{S}(\tilde\omega)=\frac{1}{2\pi}\int_{-\infty}^{\infty}e^{-i\tilde\omega\tau}
\meansb{\mf{\hat{E}}^{(-)}(\mf{R},t+\tau)\mf{\hat{E}}^{(+)}(\mf{R},t)} \ix d\tau\,.
\ee
In the far field zone the positive frequency part can be calculated as~\cite{book-agarwal}
\begin{align}
&\mf{\hat{E}}^{(+)}(\mf{R},t)=   \mf{\hat{E}}_{\text{free}}^{(+)}(\mf{R},t) \notag\\
   &-  \frac{1}{4\pi\eps_0 R c^2}  \, \sum\limits_{\mu=1}^2 \sum\limits_{i=1}^4\omega_i^2\,
\cross{\dm{i}{}}{\mf{\hat{R}}}^*\,\sSm{i}{\mu} (\hat{t}) \,e^{-ik_i\mf{\hat{R}}\cdot\mf r_\mu}.
\label{field_op}
\end{align}
The negative frequency part $\mf{\hat{E}}^{(-)}(\mf{R},t)$ can be found by Hermitian conjugation of $\mf{\hat{E}}^{(+)}(\mf{R},t)$. 
Here, $\mf{R} = R \mf{\hat{R}}$ denotes the position of the photon detector and $\hat{t}=t-R/c$ the retarded time with $c$ the speed of light. We have also introduced the cross product factor $\cross{\dm{i}{}}{\mf{\hat{R}}}$ which describes the  polarization structure of the emitted light, 
\be
\cross{\dm{i}{}}{\mf{\hat{R}}}=\mf{\hat{R}}\times (\mf{\hat{R}}\times \mf{d}_i \,)\,.
\label{cross}
\ee
In Eq.~(\ref{field_op}), the first term denotes the positive frequency part of the free field. If the point of observation lies outside the driving laser beam, it does not contribute to the correlation function.

Since we assume a polarization sensitive detector, it can detect photons emitted by $\sigma$ and $\pi$ transitions separately. 
For this, we choose the observation direction for the $\pi$ light in the $x$-$y$ plane as $\mf{\hat{R}}_\pi=(1,1,0)^T/\sqrt{2}$.
Then, products of the form $\cross{\dm{i}{}}{\mf{\hat{R}}_\pi}\crosscon{\dm{j}{}}{\mf{\hat{R}}_\pi}$ with $i\in\{1,2\}$  and $j\in\{3,4\}$ vanish. Thus we can observe the photons emitted by the $\pi$ transitions separately.

Ignoring retardation effects~\cite{milonni}, the spectrum for the linearly and circularly polarized light can be expressed as
\begin{align}
\label{spectra}
\text{S}^\pi(\tilde\omega)&=\frac{\Phi_\pi}{\pi}\int_{0}^{\infty}e^{-i\tilde\omega\tau}
\sum\limits_{\mu,\nu=1}^2\sum\limits_{i,j=1}^2\,(-1)^{i+j}\nonumber\\
&\times\means{\sSp{i}{\mu}(t+\tau)\sSm{j}{\nu}(t)}
e^{ik_0\mf{\hat{R}}_\pi(\mf{r}_\mu-\mf{r}_\nu)}\ix d\tau
\end{align}
Here, $\Phi_\pi=\omega_0^4/(4\pi\eps_0 R c^2)^2 \,\cross{\dm{1}{}}{\mf{\hat{R}}_\pi}\crosscon{\dm{1}{}}{\mf{\hat{R}}_\pi}$.  Throughout our analysis, we keep the positions of the detectors fixed, such that $\Phi_\pi$ is a constant prefactor which we will neglect in our numerical calculations.

Next, we decompose the transition operators in mean values and fluctuations, 
\be
\sSpm{i}{\mu}(t)  =  \means{\sSpm{i}{\mu}(t)}\mathds{1} \,+\, \delta \sSpm{i}{\mu}(t) \:,
\label{div_op} 
\ee
where $\mathds{1}=\mathds{1}_1\otimes\mathds{1}_2$ is the product of the two spaces belonging to the two atoms. Inserting this in Eqs.~(\ref{spectra}) it turns out that the mean value contributions lead to the coherent part of the spectra whereas the fluctuations determine the incoherent part~\cite{zubairy:qo}.
The coherent parts of the spectra evaluate to
\ba 
\text{S}^\pi_\text{coh}(\tilde\omega)&=&\delta(\tilde\omega)\,
\I{coh}{\pi}\:,
\ea
where $\I{coh}{\pi}$ denotes the total coherent resonance fluorescence intensity.

Since the coherent part of the spectrum only consists of a delta-peaked contribution at the driving laser field frequency, throughout our analysis we will focus on the incoherent part of the resonance fluorescence spectrum. In order to calculate the incoherent spectra we now define functions containing the Fourier transform of two-time averages as 
\begin{align}
&\spec{i}{j}{\mu}{\nu}{\tilde{\omega}}=\notag\\
&\int_{0}^{\infty}e^{-i\tilde\omega\tau}
\means{\delta\sSp{i}{\mu}(t+\tau)\delta\sSm{j}{\nu}(t)}\geo{\mu}{\nu}\ix d\tau\,.
\label{Tf}
\end{align}
The $\spec{i}{j}{\mu}{\nu}{\tilde{\omega}}$ can be evaluated using the quantum regression theorem~\cite{lax,carm}.
Using these functions, the incoherent $\pi$ spectrum can be written as 
\be 
\text S_\text{inc}^\pi(\tilde\omega)=\frac{\Phi_\pi}{\pi}\sum\limits_{\mu,\nu=1}^2\sum
\limits_{i,j=1}^2(-1)^{i+j}\,\spec{i}{j}{\mu}{\nu}{\tilde\omega}\:.
\label{pi1}
\ee

\subsection{\label{evalues}Eigenstate analysis}
It is well-known that the interpretation of the dynamical properties of a given quantum optical system often is facilitated by the introduction of the system dressed states. These dressed states are commonly defined as the eigenstates of the interaction picture Hamiltonian~\cite{zubairy:qo}. For our purposes, we follow a similar strategy, but include the incoherent parts of the master equation as well, since they are crucial for the SPVC and TDPC effects studied here. We start by writing the master equation Eq.~(\ref{Hg}) as a vector-matrix equation
\be 
\del{t}\wtro(t)=\MM \: \wtro(t)\:.
\ee
Here, $\wtro(t)$ is a vector containing the 256 different density matrix elements describing our system, and $\MM$ is a 256$\times$256 matrix describing the dynamics of the system. Next, we apply a unity transformation $\mc U$
\ba
\MM_{diag}&=&\mc U \MM \mc U^{\text T}\,,\\
\dwtro&=&\mc U \wtro\,
\ea
leading to a diagonal matrix $\MM_{diag}$ and a transformed state vector $\dwtro$.
The solution of the transformed master equation is given by
\be 
\dwtro(t)=e^{\,\MM_{diag} t}\:\dwtro(0)\,.
\label{diffeq}
\ee 
We express the complex eigenvalues of $\MM_{diag}$ as
\be
\xi_j=\chi_j+i\upsilon_j\:,
\ee
with $j\in\{1,\ldots,256\}$ and $\chi_j,\upsilon_j  \in\mathbb R$.
Then, the components of the solution for the transformed density matrix vector can be written as
\be 
[\dwtro(t)]_j=e^{\xi_j t}\;[\dwtro(0)]_j=e^{(\chi_j+i\upsilon_j)t}\;[\dwtro(0)]_j\:.
\ee
Similar to the dressed state analysis, we find that negative real parts of the eigenvalues lead to an exponential decay of  $\dwtro(t)$, whereas the imaginary parts result in an energy shift. Therefore, the $\upsilon_j$ can be used to understand the origin of the different lines in the resonance fluorescence spectrum.

\begin{figure}
\includegraphics[width=8cm]{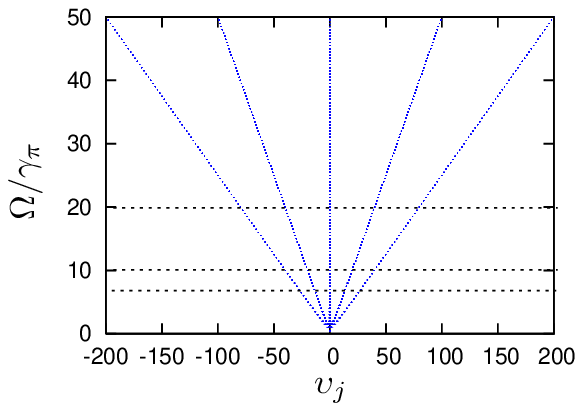}
\includegraphics[width=8cm]{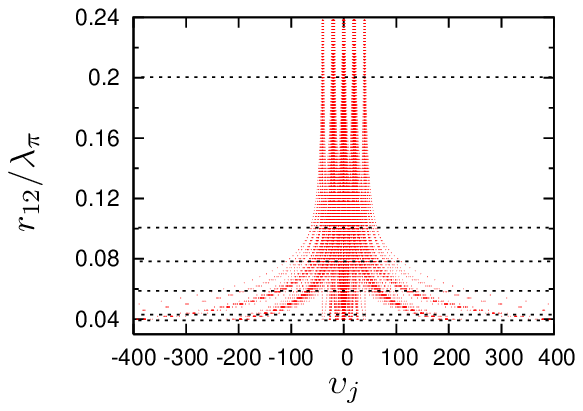}
\caption{Energy shifts of the system dressed states, evaluated as the imaginary part $\ups_j$ of the eigenvalues $\xi_j=\chi_j+i\upsilon_j$ of the matrix $\MM$ governing the system dynamics. The interatomic distance vector has direction given by $\theta=\pi/2$ and $\phi=\pi/4$. (a) shows the case of large interatomic distance $r_{12}=10\lambda_\pi$ in dependence of the Rabi frequency $\Omega$. The four branches clearly show the splitting of the spectrum into a central feature and Mollow sidebands due to the AC-stark splitting of the bare states. The horizontal dotted lines mark the values for $\Omega$ chosen for calculations in the later sections. (b) depicts the energy shifts as a function of the interatomic distance for fixed Rabi frequency $\Omega=10\gamma_\pi$. The three different cases of large, intermediate and small distance can clearly be distinguished. The horizontal dotted lines mark distances for which we will present resonance fluorescence spectra in later sections.}
\label{OmEW}
\end{figure}

\section{\label{sec-result}Results}

\subsection{\label{sec-eigenstate}Eigenstate analysis}
We start by analyzing the position of the system's dressed states in order to identify different parameter ranges of interest. Fig.~\ref{OmEW}(a) shows the imaginary parts $\upsilon_j$ of all eigenvalues of the matrix $\MM_{diag}$ in dependence of the Rabi frequency and at a large interatomic distance $r_{\mu\nu}=r_{12}=10\lambda_\pi$. In this case, the two atoms essentially act independently. Here, the imaginary parts of the eigenvalues lie around $0,\pm\Omega/\gamma_\pi$ and $\pm 2\Omega/\gamma_\pi$. It should be noted, however, that not all potential positions indicated by the eigenstate analysis do lead to a significant peak in the spectrum. In fact, in \cite{martin1} was shown that in the spectrum of a single such a four level atom only peaks at $0,\pm\Omega/\gamma_\pi$ occur. This Mollow spectrum we also have to expect for two independent atoms.
In Fig.~\ref{OmEW}(b), the positions of the eigenstates are shown against the interatomic distance $r_{12}$ for fixed Rabi frequency $\Omega=10\gamma_\pi$. Three different distance regimes can be distinguished. In the large-distance case which is already reached at $r_{12} \gtrsim 0.2 \lambda_\pi$, the eigenvalue spectrum is dominated by the Mollow structure already shown in Fig.~\ref{OmEW}(a). In the small-distance case $r_{12} \lesssim 0.05 \lambda_\pi$, different eigenvalue branches with positions depending on the interatomic distance are formed. These can be attributed to the TPVC-induced energy shifts, with a $r_{12}^{-3}$-dependence of the interatomic coupling constants. Finally, in the intermediate distance case $0.05 \lambda_\pi \lesssim r_{12} \lesssim 0.2 \lambda_\pi$, the eigenvalues overlap such that a clear interpretation becomes difficult.

Next, we will present results for the spectra in these three different regimes of interatomic distances, and in detail explain the structure of the obtained spectra based on the different coupling mechanisms.

\subsection{\label{sec-inter}Interatomic couplings}
In the following we study the influence of the interatomic couplings on the resonance fluorescence spectra. Since these parameters are negligible for large $r_{12}$ we mostly choose smaller distances as compared to the wavelength $\lambda_\pi$. We will see that additional peaks in the spectra occur due to these couplings between the two atoms as observed when calculating the eigenvalues of the matrix $\MM$ in Sec.~\ref{evalues}. Artificially turning on and off some coupling constants will help us to get a better understanding of the peak positions and thus the relevant physical coupling mechanisms. We assume our atoms to be located in the $x$-$y$ plane which means  $\theta=\pi/2$ and we choose $\phi=\pi/4$. Therefore, the couplings between one $\pi$ and one $\sigma$ transition vanish and thus no couplings between one $\pi$ and one $\sigma$ dipole can occur. 
\subsubsection{\label{interm}Spectra for large and intermediate interatomic distances}

\begin{figure}
\includegraphics[width=8cm]{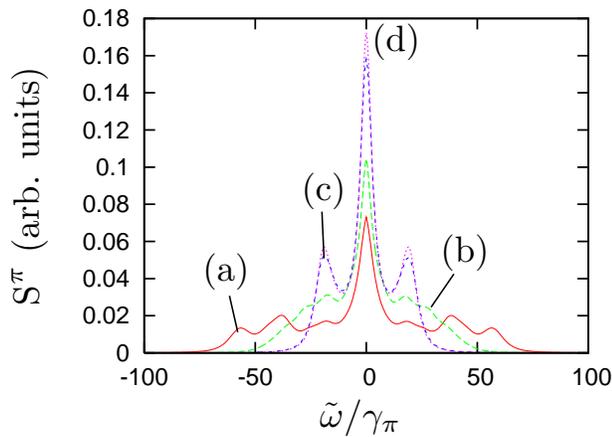}
\caption{Linearly polarized resonance fluorescence spectrum for different interatomic distances $r_{12}$. In curve (a) we chose $r_{12}=0.08\lambda_\pi$, in (b) $r_{12}=0.1\lambda_\pi$, in (c) $r_{12}=0.2\lambda_\pi$ and in (d) $r_{12}=10\lambda_\pi$. In all curves the Rabi frequency is $\Omega=10\gamma_\pi$, and the laser field is applied on resonance $\Delta=0$. The interatomic distance vector is aligned with $\theta=\pi/2$ and $\phi=\pi/4$.}
\label{rgross}
\end{figure}

We choose the Rabi frequency as $\Omega=10\gamma_\pi$. 

In Fig.~\ref{rgross}, we show the linearly polarized spectrum observed from the $(1,1,0)$ direction. As interatomic distances we choose $r_{12}=0.08\lambda_\pi$ in curve (a), $r_{12}=0.1\lambda_\pi$ in (b), $r_{12}=0.2\lambda_\pi$ in (c), and the large distance $r_{12}=10\lambda_\pi$ in (d). 
Since the interatomic couplings vanish for large distances, in this regime the atoms behave like two independent particles. Then for the $\pi$ spectrum we obtain one peak at the laser frequency where $\tilde\omega=\omega-\omega_L=0$ and two sideband peaks at $\pm 2\Omega$ as in the Mollow spectrum, see curve (d) in the upper subfigure. %
The curve for $r_{12}=0.2\lambda_\pi$ is still similar to that for large distances. Only the amplitudes of the peaks differs from the Mollow spectrum for large distances. Thus the interatomic couplings can be considered weak for a distance as small as $0.2\lambda_\pi$, which is in agreement with our findings from Fig.~\ref{OmEW}.
If we further decrease $r_{12}$ the peaks at $\tilde\omega=0$ and $\tilde\omega=\pm 2\Omega$ become lower and additional peaks arise at higher frequencies $|\tilde\omega|$, see curves (a) and (b).  In Fig.~\ref{OmEW} we marked the values $r_{12}=0.08\lambda_\pi$ and $0.1\lambda_\pi$ by dotted lines. This regime of interatomic distances gives rise to rather complicated dynamics, and several $\ups_j$ occur very close to each other.  This is also reflected in the spectra, which is characterized by overlapping peaks. The reason for this is that  the interatomic coupling parameters are of the same order of magnitude as the laser Rabi frequency $\Omega$ in this intermediate distance regime. 

\subsubsection{\label{small}Spectra for small interatomic distances}

\begin{figure}
\includegraphics[width=8cm]{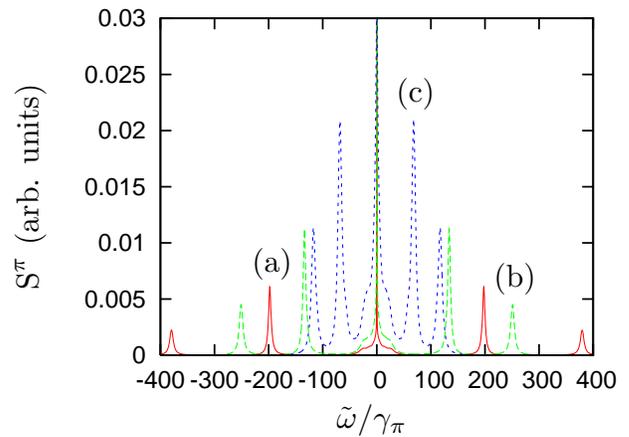}
\caption{\label{rklein}Resonance fluorescence spectrum at small interatomic distances. In (a) $r_{12}=0.04\lambda_\pi$, in (b) $r_{12}=0.046\lambda_\pi$ and in (c) $r_{12}=0.06\lambda_\pi$. All other parameters are chosen as in Fig.~\ref{rgross}.}
\end{figure}

In Fig.~\ref{rklein}, we show results for interatomic distances chosen as   (a) $r_{12}=0.04\lambda_\pi$, (b) $r_{12}=0.046\lambda_\pi$, and (c) $r_{12}=0.06\lambda_\pi$. In this regime of small distances we can see four clear sideband peaks in the spectrum instead of two in the Mollow spectrum. Since these sideband peaks move farther away from the center $\tilde\omega=0$ when decreasing $r_{12}$, they can be associated to the TPVC which increase in magnitude with decreasing distance. For the linearly polarized light we still obtain a peak at $\tilde\omega=0$ which becomes higher and narrower with decreasing interatomic distances.
These spectra again correspond to the results of Fig.~\ref{OmEW}. One can see that the accumulation of imaginary parts of eigenvalues $\ups_j$ for intermediate $r_{12}$ splits up into clear branches when decreasing the interatomic distance.

\subsubsection{\label{ident}Interpretation of the spectra}

\begin{figure}
\includegraphics[width=8cm]{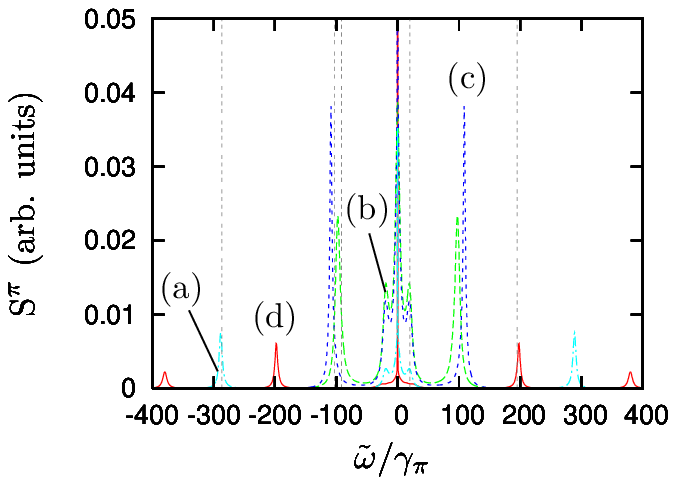}
\vspace{0.2cm}
\includegraphics[width=8cm]{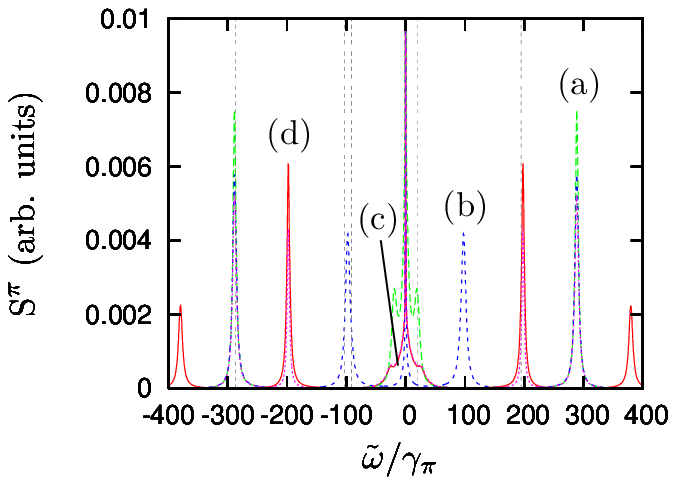}
\caption{Here, we show the $\pi$ spectrum for the distance $r_{12}=0.04\lambda_\pi$ while we artificially set some interatomic couplings equal to zero. In each curve of the upper subfigure except curve (d) only the couplings of one $\mf{\mc G}_a$ are on. In curve (a) the parameters of $\mf{\mc G}_2$ are turned on, in (b) the ones of $\mf{\mc G}_3$, in (c) the ones of $\mf{\mc G}_3$ and in (d) we plot the spectrum with all couplings on. In the lower subfigure we turn one group of interatomic couplings after the other on. In curve (a) only coupling parameters of $\mf{\mc G}_2$ are on, in (b) the couplings of $\mf{\mc G}_2$ and $\mf{\mc G}_3$, in (c) the constants of $\mf{\mc G}_2$, $\mf{\mc G}_3$ and $\mf{\mc G}_4$ are on and curve (d) depicts the spectrum with all couplings on. All other parameters are chosen as in Fig.~\ref{rgross}.}
\label{pikopplaus}
\end{figure}

In this Section we proceed to interpret the origin of the different spectral features found in the previous section. In order to do so, we artificially switch off parts of the interatomic couplings, and observe the change in the corresponding spectra. Note that in the following discussion, the SPVC are always kept in the analysis. 
For this procedure we divide our interatomic coupling constants into five groups $\mf{\mc G}_1,\; \mf{\mc G}_2,\;\mf{\mc G}_3,\;\mf{\mc G}_4$ and $\mf{\mc G}_5$.
These groups are defined as follows
\begin{align}
\mf{\mc G}_1&=\left\lbrace\Gamma_{ij}^{\mu\nu} , \Omega_{ij}^{\mu\nu}|\: i\in\{1,2\} \ix{and }j\in\{3,4\}\right\rbrace
\nonumber\\
&\hspace{2cm} \cup \left\lbrace\Gamma_{ij}^{\mu\nu} , \Omega_{ij}^{\mu\nu}|\: i\in\{3,4\} \ix{and }j\in\{1,2\}\right\rbrace\nonumber\\
&=\left\lbrace\ix{couplings between one $\pi$ and one $\sigma$ dipole }\right\rbrace\nonumber\\
\mf{\mc G}_2&=\left\lbrace\Gamma_{ij}^{\mu\nu} , \Omega_{ij}^{\mu\nu}|\: i,j\in\{3,4\}\ix{and } i\neq j\right\rbrace\nonumber\\
&=\lb\ix{couplings between two different $\sigma$ dipoles }\right\rbrace\nonumber\\
\mf{\mc G}_3&=\left\lbrace\Gamma_{ij}^{\mu\nu} , \Omega_{ij}^{\mu\nu}|\:\mu\neq\nu\ix{and } i,j\in\{1,2\}\ix{and } i\neq j\right\rbrace\nonumber\\
&=\lb\ix{interatomic couplings between}\right. \nonumber\\
&\hspace{2cm} \left.\ix{two different $\pi$ dipoles }\right\rbrace\nonumber\\
\mf{\mc G}_4&=\left\lbrace\Gamma_{ij}^{\mu\nu} , \Omega_{ij}^{\mu\nu}|\:\mu\neq\nu\ix{and } i,j\in\{3,4\}\ix{and } i=j\right\rbrace\nonumber\\
&=\lb\ix{interatomic couplings between} \right.\nonumber\\
& \hspace{2cm} \left.\ix{two equal $\sigma$ dipoles }\right\rbrace\nonumber\\
\mf{\mc G}_5&=\left\lbrace\Gamma_{ij}^{\mu\nu} , \Omega_{ij}^{\mu\nu}|\:\mu\neq\nu\ix{and } i,j\in\{1,2\}\ix{and } i=j\right\rbrace\nonumber\\
&=\lb\ix{interatomic couplings between}\right. \nonumber\\
& \left.\hspace{2cm}\ix{two equal $\pi$ dipoles }\right\rbrace\:.
\label{groups}
\end{align}
$\g{1}$ and $\g{2}$ also contain some intraatomic couplings. However, these are zero because of the orthogonality of the respective dipole moments. Additionally, if both atoms are in the $x$-$y$ plane, which means $\theta=\pi/2$, all interatomic couplings of group $\g{1}$ vanish.
Since our system contains only two atoms and the couplings do not change when exchanging $\mu$ and $\nu$ we use as superscript of the interatomic couplings $inter$ instead of $\mu\nu$.
Our classification is done in such a way, that within one group the absolute values of both real and imaginary part of all $\Gamma_{ij}^{inter}$ or $\Omega_{ij}^{inter}$, respectively, are equal. This still holds if we choose $\theta\neq \frac{\pi}{2}$. 
For each group $\g{a}$ we define one pair of coupling parameters ($\Gamma_a, \Omega_a)$ as representative for this group. These definitions are made as follows
\ba
\Gamma_1&=\Gamma_{13}^{inter}\:,\hspace{2cm}\Omega_1&=\Omega_{13}^{inter}\:,\notag\\
\Gamma_2&=\Gamma_{34}^{inter}\:,\hspace{2cm}\Omega_2&=\Omega_{34}^{inter}\:,\notag\\
\Gamma_3&=\Gamma_{12}^{inter}\:,\hspace{2cm}\Omega_3&=\Omega_{12}^{inter}\:,\notag\\
\Gamma_4&=\Gamma_{33}^{inter}\:,\hspace{2cm}\Omega_4&=\Omega_{33}^{inter}\:,\notag\\
\Gamma_5&=\Gamma_{11}^{inter}\:,\hspace{2cm}\Omega_5&=\Omega_{11}^{inter}\:.
\ea
The absolute values $|\Gamma_{ij}^{inter}|$ and $|\Omega_{ij}^{inter}|$ of all pairs of coupling constants belonging to the same group are equal.

\begin{figure}
\includegraphics[width=8cm]{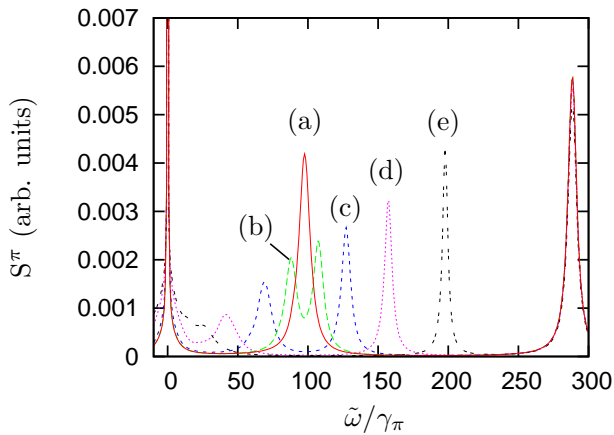}
\caption{Incoherent resonance fluorescence spectrum emitted by the $\pi$ transitions for the case where the couplings of $\g{2}$ and $\g{3}$ are considered and in addition we progressively switch on the parameters of $\g{4}$. In curve (a) the coupling constants of $\g{4}$ are zero, in (b) they are multiplied by $0.1$, in (c) by $0.3$, in (d) by $0.6$ and in (e) by one. All other parameters are as in Fig.~\ref{pikopplaus}.}
\label{G4}
\end{figure}

In Fig.~\ref{pikopplaus}, our results for the spectrum emitted by the $\pi$ transitions are shown. We choose $r_{12}=0.04\lambda_\pi$ and the Rabi frequency $\Omega=10\gamma_\pi$. Since our atoms are located in the $x$-$y$ plane, all couplings of $\mf{\mc G}_1$ vanish. The upper subfigure shows spectra obtained when only one group of couplings is on and the others are set equal to zero.  In curve (a) all parameters of $\mf{\mc G}_2$, in (b) the ones of $\mf{\mc G}_3$, and in (c) the ones of $\mf{\mc G}_4$ are on. Turning only coupling constants of $\mf{\mc G}_5$ on gives us the same spectrum as in curve (b). Thus we conclude that the interatomic interaction between equal $\pi$ dipoles and between two different ones has similar impact on the dynamics. This is not surprising since $\dm{1}{}$ and $\dm{2}{}$ are antiparallel. 

All curves show peaks at $\tilde\omega=0$ and $\tilde\omega=\pm 2\Omega$ due to the driving laser field as in the spectrum without interatomic couplings, see Fig.~\ref{rgross}.
In the spectrum plotted in curve (a) we can see additional peaks at $\tilde\omega\approx\pm|\Omega_2|=\pm 286.5\gamma_\pi$. These peaks arise due to an energy splitting of the atomic levels caused by the vacuum-mediated coupling of the atoms. Curve (b) shows additional maxima at approximately $\tilde\omega\approx\pm|\Omega_3|=\pm 91.6\gamma_\pi$. Looking closely one can see that the peak frequencies are a little bit higher than the coupling parameter. The reason for this likely is the  influence of the Rabi frequency on the position of these outer peaks. This also holds for other peaks analyzed below. 
In curve (c) peaks at $\tilde\omega\approx\pm 103.2\gamma_\pi$ occur which is equal to $\pm|\Omega_4|$.

Now we continue by turning on the coupling constants one after the other. In the lower subfigure of Fig.~\ref{pikopplaus} the resulting spectra are depicted. Curve (a) is the same as in the upper subfigure (only the plot range is changed).

Curve (b) shows the $\pi$ spectrum where the couplings of $\g{2}$ and $\g{3}$ differ from zero. We can see four peaks in addition to the triplet around $\tilde\omega=0$. These new peaks are located at approximately $\tilde\omega\approx\pm|\Omega_2|=\pm 286.5\gamma_\pi$ and $\tilde\omega\approx\pm |\Omega_3|=\pm 91.6\gamma_\pi$. Note that the maxima around $\pm 286.5\gamma_\pi$ are hardly visible in the graph, since other curves show peaks at the same positions.

\begin{figure}
\includegraphics[width=8cm]{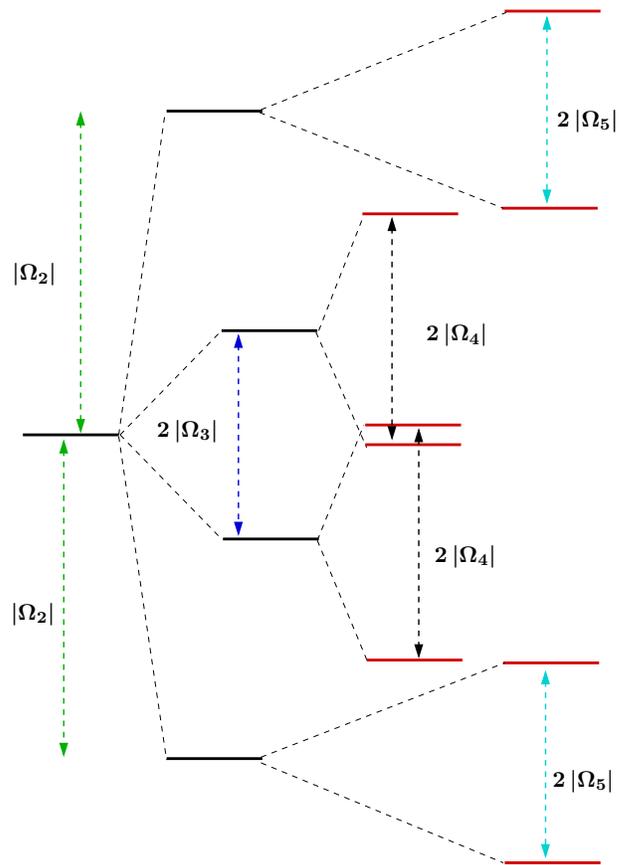}
\caption{Splitting of the atomic energy levels due to the interatomic coupling parameters for a small distance $r_{12}=0.04\lambda_\pi$. For this interatomic distance the values of the coupling constants are $|\Omega_2|=286.5\gamma_\pi$, $|\Omega_3|=|\Omega_5|=91.6\gamma_\pi$ and $|\Omega_4|=103.2\gamma_\pi$.}
\label{levauf}
\end{figure}

In order to obtain curve (c) we turn the couplings of $\g{4}$ on in addition to $\g{2}$ and $\g{3}$. Then the peaks at about $\tilde\omega\approx\pm |\Omega_3|$ vanish and new maxima around $\tilde\omega=\pm 200\gamma_\pi$ arise. Since this is not equal to the value of any coupling parameter, this peak cannot be interpreted straightforward as the peaks in curve (a) and (b). In order to find the origin of these peaks around $\tilde\omega=\pm 200\gamma_\pi$ we turn the couplings of $\g{4}$ progressively on, see Fig.~\ref{G4}. Here, progressively  means that we multiply the respective coupling parameters by a factor $p\in[0,1]$. For curve (a) the couplings of $\g{4}$ are set zero ($p=0$), in (b) $p=0.1$, in (c) $p=0.3$, in (d) $p=0.6$, in (e) $p=0.9$, and in (f) the respective couplings are on which means $p=1$. We can observe that the peaks at about $\pm |\Omega_3|=\pm 91.6\gamma_\pi$ split up into two peaks. This splitting becomes larger for increasing values of the factor $p$. For $p=1$ where the couplings of $\g{4}$ are on, the respective peaks are located at $\tilde\omega\approx\pm 11\gamma_\pi=|\Omega_3|-|\Omega_4|$ and $\tilde\omega\approx 194.8\gamma_\pi=|\Omega_3|+|\Omega_4|$. Note that the maxima at about $\pm 11\gamma_\pi$ cannot be distinguished clearly from the ones caused by the Rabi frequency of the driving laser field at $\pm 20\gamma_\pi$ and the centered peak at $\tilde\omega=0$. During this level splitting when slowly turning on the couplings of $\g{4}$ the peaks at about $\pm 286.5\gamma_\pi=\pm |\Omega_2|$ do not change. This means that the couplings between equal $\sigma$ dipoles ($\Omega_4$) and between different $\sigma$ dipoles ($\Omega_2$) enter the dynamics of our system independently. By contrast, the level splitting of Fig.~\ref{G4} shows us that the couplings between $\pi$ dipoles and between equal $\sigma$ dipoles influence each other.

Additionally turning on the parameters of $\g{5}$ we obtain the spectrum with all interatomic couplings as already plotted in the upper subfigure of Fig.~\ref{pikopplaus}. Here, the peak at $\pm|\Omega_2|=\pm 286.5\gamma_\pi$ no longer exists but is replaced by maxima at $\tilde\omega\approx\pm(|\Omega_2|\pm|\Omega_5|)=\pm(286.5\gamma_\pi\pm 91.6\gamma_\pi)$. Note that for our parameters $|\Omega_3|+|\Omega_4|=91.6\gamma_\pi+103.2\gamma_\pi=194.8\gamma_\pi\approx|\Omega_2|
-|\Omega_5|=286.5\gamma_\pi-91.6\gamma_\pi=195.5\gamma_\pi$ and therefore the peaks at these positions cannot be distinguished. To confirm the splitting of the peaks at $\tilde\omega\approx 286.5\gamma_\pi$ in curve (c) we do the same procedure for $\g{5}$ as done for $\g{4}$. We turn all couplings of $\g{5}$ progressively on while the parameters of $\g{2}$, $\g{3}$ and $\g{4}$ are always on. Here we found a similar peak splitting as for $\g{4}$. 

Finally, we conclude our findings in this section by summarizing the various splittings in a dressed state analysis shown in Fig.~\ref{levauf}.

\subsection{\label{sec-intra}Intraatomic couplings}

\subsubsection{Classification of intraatomic couplings}
\begin{figure}
\includegraphics[width=8cm]{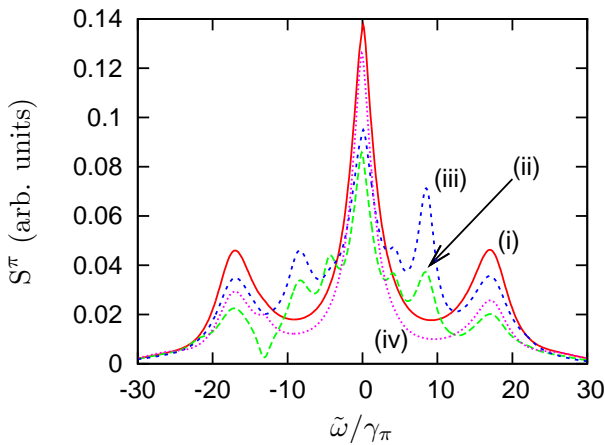}
\caption{\label{intra-1}Impact of the intraatomic couplings on the resonance fluorescence spectrum emitted by the $\pi$ transitions for intermediate distances. The parameters are $r_{12}=0.09\lambda_\pi$, $\Omega=6\gamma_\pi$, $\Delta = -14\gamma_\pi$, $\theta=\pi/2$ and $\phi=\frac{\pi}{4}$. In curve (i) the complete spectrum with all couplings is shown as it could be observed in an experiment. In (ii), all intraatomic couplings are artificially turned off.  In (iii), all intraatomic couplings entering the equations of motion are kept, while those entering the expression for the spectrum Eq.~(\ref{spectra}) are set to zero. In (iv), the intraatomic couplings in the expression for the spectrum are kept while those in the equations of motion are dropped.}
\end{figure}
We now turn to the intraatomic (SPVC) couplings.
In our system we can distinguish between two different types of intraatomic couplings. The first type are the coupling constants entering in the equations of motion for our system, $\Gamma_{ij}^{\mu\mu}$ and $\Omega_{ij}^{\mu\mu}$ (where $\Omega_{ij}^{\mu\mu}$ are always zero)  with $i\neq j$.  The second type are contributions of second-order correlation functions to the spectrum in Eq.~(\ref{pi1}) where $\mu=\nu$. We are thus led to split the resonance fluorescence spectrum into four contributions,
\begin{align}
\text S_\text{inc}^\pi(\tilde\omega)&= \mc P_1(\tilde\omega)+\mc P_2(\tilde\omega)-\mc P_3(\tilde\omega)-\mc P_4(\tilde\omega)\,,
\end{align}
where
\begin{subequations}
\begin{align}
\mc P_1(\tilde\omega)&=\frac{\Phi_\pi}{\pi}\sum\limits_{\mu=1}^2\sum
\limits_{i=1}^2\spec{i}{i}{\mu}{\mu}{\tilde\omega}\,,\\
\mc P_2(\tilde\omega)&=\frac{\Phi_\pi}{\pi}\sum\limits_{\substack{\mu,\nu=1\\ \mu\neq\nu}}^2\sum
\limits_{i=1}^2\spec{i}{i}{\mu}{\nu}{\tilde\omega}\,,\\
\mc P_3(\tilde\omega)&=\frac{\Phi_\pi}{\pi}\sum\limits_{\mu=1}^2\sum
\limits_{\substack{i,j=1\\ i\neq j}}^2\spec{i}{j}{\mu}{\nu}{\tilde\omega}\\
\mc P_4(\tilde\omega)&=\frac{\Phi_\pi}{\pi}\sum\limits_{\substack{\mu,\nu=1\\ \mu\neq \nu}}^2\sum
\limits_{\substack{i,j=1\\ i\neq j}}^2\spec{i}{j}{\mu}{\nu}{\tilde\omega}\,.
\end{align}
\end{subequations}
We then denote $\mc P_1-\mc P_3$ as the intraatomic part of the contributions to the spectrum, and  $\mc P_1+\mc P_2-\mc P_4$ as the interatomic contribution. It should be noted, however, that this separation is difficult in particular due to the term $\pm \mc P_4$. This term contributes for $\mu\neq \nu$ which indicates an interatomic contribution, but at the same time it only contributes for $i\neq j$, which is the characteristic for an intraatomic coupling. One consequence of this is that spectra which are calculated by artificially suppressing e.g., the intraatomic parts, are not necessarily positive definite, which makes a straightforward interpretation difficult. It is important to keep in mind that results obtained by artificially suppressing parts of the couplings do not correspond to physically observable situations, but can only be used to interpret the obtained full spectra in certain situations, as explained in more detail below.

\begin{figure}
\includegraphics[width=8cm]{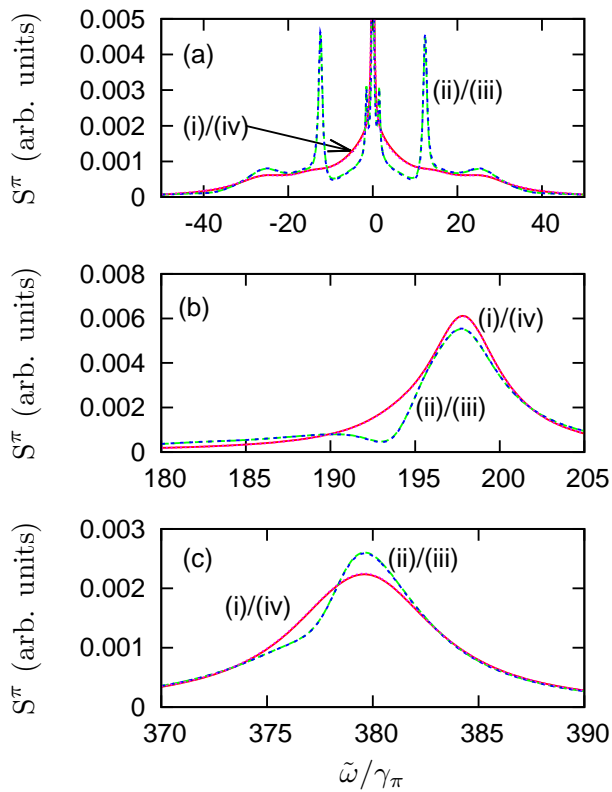}

\caption{\label{intra-2}Impact of the intraatomic couplings on the resonance fluorescence spectrum emitted by the $\pi$ transitions in the small distance case. The parameters are as in Fig.~\ref{rklein}(a). The three subfigures separately show the three spectral features visible in the right half of the spectrum in Fig.~\ref{rklein}(a). Each subfigure shows four curves (i)-(iv). As in Fig.~\ref{intra-1}, these correspond to the complete spectrum (i), the spectrum without intraatomic couplings (ii), and the spectrum with either the intraatomic coupling entering the expression for the spectrum artificially suppressed (iii) or those entering the equations of motion switched off (iv).}
\end{figure}

\subsubsection{Effect of the intraatomic couplings on the resonance fluorescence}
We start our discussion by noting that in principle, both the intraatomic couplings in the equations of motion and in the expression for the spectrum can crucially modify the total observed spectrum. This is surprising, since  these couplings in the equations of motion do not contribute to the steady state, and thus the resonance fluorescence spectrum, of the corresponding single atom system~\cite{martin1}. The relevance of both types of  couplings  can most readily be seen for parameter ranges in which the intraatomic coupling constants $\Gamma_{ij}^{\mu\mu}$ are of similar magnitude as the interatomic coupling constants $\Omega_{ij}^{\mu\nu}$ and $\Gamma_{ij}^{\mu\nu}$ ($\mu\neq\nu$). An example is shown in Fig.~\ref{intra-1}. The figure shows the total resonance fluorescence spectrum (i) emitted by the $\pi$ transitions, as well as the corresponding spectra obtained by artificially switching off the intra-atomic couplings in the equations of motion (iii), in the spectrum Eq.~(\ref{spectra}) (iv), or both (ii). It can be seen that all four curves differ considerably, and we conclude that both types of intra coupling are of relevance. It is, however, difficult to attribute certain features of the observable total spectrum to either of the two contributions. The reason is that for the parameters in Fig.~\ref{intra-1}, 
the total spectrum cannot simply be decomposed into the spectrum without intraatomic couplings and the two corrections arising from the couplings in the equations of motion and in the expression for the spectrum.

We now turn to smaller interatomic distances in order to study the role of the SPVC in the case of numerically dominating TPVC. Since the interatomic coupling constants then are much larger than the intraatomic coupling constants, typically the intraatomic coupling constants $\Gamma_{ij}^{\mu\mu}$ entering the equations of motion only slightly influence the total spectrum. 
An example is shown in Fig.~\ref{intra-2}, for parameters as in Fig.~\ref{rklein}(a). The three subfigures show the three spectral features in the ranges $[-50,50]$, $[180,205]$ and $[370,390]$ of the right hand part of the spectrum visible in Fig.~\ref{rklein}(a). Again, the total spectrum, as well as the corresponding spectrum with one or both of the intraatomic couplings artificially suppressed are shown. It can be seen that despite the clearly numerically dominating interatomic coupling constants, the spectrum is strongly modified upon suppression of the intraatomic coupling. In contrast to the example in Fig.~\ref{intra-1}, however, a grouping into two pairs of curves is observed. The curve for the full spectrum (i) coincides with the spectrum  (iii) with intraatomic couplings entering the equations of motion artificially suppressed. Similarly, the spectrum (ii) without intraatomic coupling coincides with the result (iv) obtained by suppressing the intraatomic couplings in the expression for the spectrum. From this, we conclude that for these parameters, the influence of the intraatomic coupling clearly arises from the corresponding parts in the expression for the spectrum, whereas the direct couplings in the equations of motion only give rise to minor corrections.
This is the situation also found in the corresponding single atom system.

\subsubsection{Intraatomic couplings induced by interatomic couplings}

We found in Fig.~\ref{intra-1} that already the intraatomic couplings in the equations of motion alone can give rise to a significant modification of the resonance fluorescence spectrum, in contrast to the corresponding single-atom case. To interpret this difference in close analogy to the single atom case, as a first step, we evaluated the full steady state density matrix for parameters as in Fig.~\ref{intra-1} both for the case with all couplings on, and with intraatomic couplings in the equations of motion suppressed. It turns out that while the magnitude and phases of the entries in the two density matrices differ, no elements are zero in one of the cases and non-zero in the other case. We thus conclude that no fundamentally new coherences are created due to the intraatomic couplings in the equations of motion. Next, for a better comparison with the single atom case, we analyze the steady state density matrix for one of the two atoms in our system. For this, we calculate the stationary state of the total system and then trace out the second atom. 
\begin{figure}[t]
\includegraphics[width=8cm]{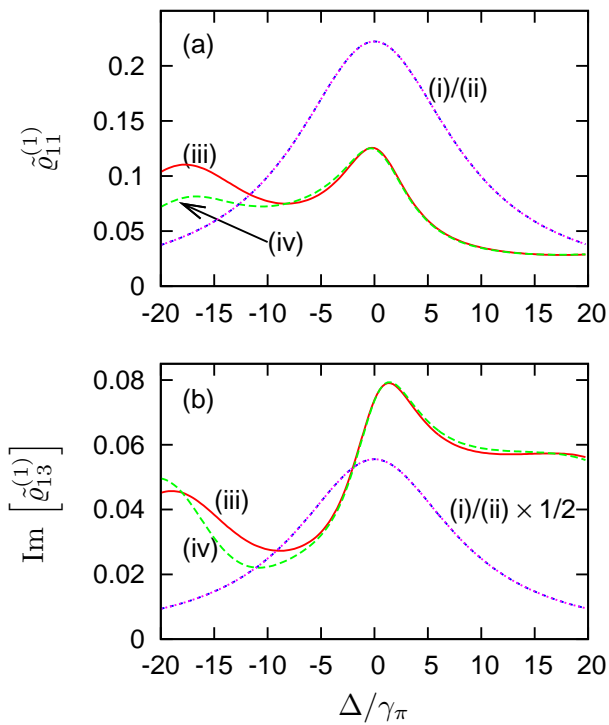}

\caption{\label{induce-1}Intraatomic couplings induced by interatomic couplings. (a) shows the steady state population of state $|1\rangle$ in atom 1 obtained by tracing over the second atom. (b) shows the imaginary part of the coherence between states $|1\rangle$ and $|3\rangle$ in atom 1. The different curves are as follows:  (i) All couplings included, large-distance case corresponding to no interatomic couplings. (ii) Large distance case, without intraatomic couplings entering the equations of motion. (iii) and (iv) show the corresponding results for small distance $r_{12} = 0.09\lambda_\pi$. The parameters are as in Fig.~\ref{intra-1} except for the variable detuning $\Delta$. Note that in (b), the two curves (i) and (ii) are shown multiplied with a factor 1/2 for better visibility.}
\end{figure}
As a first example, Fig.~\ref{induce-1}(a) shows the population of state $|1\rangle$ of atom 1 against the detuning $\Delta$. From this figure it can be seen that at large interparticle distance, the intraatomic couplings entering the equations of motion do not have any effect, as expected from the single-atom case. However, at small distances, the populations with and without these couplings differ considerably in a range of negative detunings. In this sense, the intraatomic couplings between the two particles induce an effect of the intraatomic couplings which could not be observed in a single-atom system. This mechanism of interatomic couplings inducing intraatomic couplings is responsible for the dependence of the resonance fluorescence spectra on the intraatomic couplings in the equation of motion found in Fig.~\ref{intra-1}.
In Fig.~\ref{induce-1}(b), we show corresponding results for the imaginary part of the coherence between states $|1\rangle$ and $|3\rangle$ of atom 1. We again find that the interatomic couplings induce an effect of the intraatomic couplings entering the equations of motion. 

A possible explanation for these induced couplings is that in the single-particle case, the SPVC entering the equations of motion do not contribute to the steady state of the density matrix since the two ground states $|3\rangle$ and $|4\rangle$ are orthogonal, as explained in the introduction. In contrast, in the two-particle case, the relevant states are collective eigenstates originating from the TPVC, which each consist of different bare atomic states. Between these collective eigenstates, a modified set of transitions with different dipole moments occur. In particular, due to bare state mixing, near-degenerate non-orthogonal transition pathways originating from a single collective eigenstate may be created. An analogous mechanism was suggested as a way to induce spontaneously induced coherences in a single three-level system in $\Lambda$ configuration~\cite{book-ficek,induce}. The two bare state transitions from the upper state $|e\rangle$ to the lower states $|a\rangle$ and $|b\rangle$ originally are assumed to have orthogonal transition dipole moments. If a resonant laser field is applied to transition $|a\rangle \leftrightarrow |e\rangle$, then the system can be described in the dressed state basis $\{|+\rangle, |-\rangle, |b\rangle\}$, where $|\pm\rangle = (|e\rangle \pm |a\rangle)/\sqrt{2}$. In this new basis, the two transitions from the dressed upper states $|\pm\rangle$ to the lower state $|b\rangle$ are near degenerate and non-orthogonal, such that quantum interference can take place.

We thus conclude that the impact of the intraatomic coupling constants remains important even at low interatomic distances for which the interatomic coupling constants are much larger than the corresponding intraatomic ones. For a large range of parameters, and in particular at smaller distances, the dominating contribution of the intraatomic coupling constants arises from the parts entering the expression Eq.~(\ref{spectra}) of the spectrum. In contrast to the corresponding single atom case,  in a certain parameter range, the contributions entering the equations of motion can also have a substantial influence on the obtained spectra. Since this contribution only occurs for the case of two nearby atoms, we conclude that the influence of the intraatomic couplings is induced by the intraatomic couplings.

\section{\label{sec-summary}Summary}
In summary, we have analyzed a system of two dipole-dipole interacting nearby four-level atoms in 
$J=1/2 \leftrightarrow J=1/2$ configuration. This is the simplest model system which on the one hand provides a complete description of the dipole-dipole interaction for arbitrary orientation of the interatomic distance vector, and on the other hand allows for spontaneously generated coherences in a realistic atomic level scheme. The complete description of the dipole-dipole interactions is achieved by considering full Zeeman manifolds for the ground and the excited state. The spontaneously generated coherences contribute since this level scheme has two near-degenerate dipole allowed transitions with (anti-)parallel dipole moments. However, as these two transitions do not share a common state, the comprehensive theoretical results on spontaneously generated coherences such as in the usual $V$-type or $\Lambda$-type configuration cannot be applied directly. We discuss different methods to analyze the contribution of the various coupling constants to the total resonance fluorescence spectrum. A first analysis is possible based on the eigenvalue spectrum of the matrix governing the system dynamics. Then, we artificially suppress certain groups of couplings, in order to reveal their significance for the total spectrum. We also gradually switch on selective couplings by artificially multiplying the corresponding coupling coefficient with a parameter ranging from zero to unity. This allows us to find a dressed state interpretation of the contribution of the different interatomic dipole-dipole couplings to the total spectrum. 

Regarding the intraatomic couplings, we identify two different types of contributions. The first is via the SPVC-induced coupling coefficients directly entering the equation of motion. The second contribution appears in the expression for the resonance fluorescence spectrum. In general, both contributions can substantially influence the total resonance spectrum, even though it is difficult to attribute specific spectral features to either of the two contributions. However,  in particular for smaller interatomic distances, we find that the dominant contribution of the intraatomic coupling is the one appearing in the expression for the spectrum. A simple interpretation is that for small interatomic distances, the interatomic coupling coefficients entering the equations of motion are much larger than the corresponding intraatomic ones, such that they typically have only a small contribution. Nevertheless, we find that the intraatomic couplings have an observable impact even at small interatomic distances, such that also the two-particle 
$J=1/2 \leftrightarrow J=1/2$ system is an interesting candidate to observe these intraatomic couplings. Finally, we could show that the intraatomic couplings entering the equations of motion can have a significant effect on the observed spectra, in contrast to the single-particle case. We thus conclude that the interparticle couplings can induce additional contributions of the intraatomic couplings.


\end{document}